\documentclass[referee]{aa} 
\usepackage{graphicx}
\usepackage{natbib}
\bibpunct{(}{)}{;}{a}{}{,} 
\begin{document}
\titlerunning{ISM dust feedback from low to high mass stars}
   \title{The role of stellar mass and mass functions on the ISM dust feedback}

   \author{D. Falceta-Gon\c calves
          \inst{} 
		}

   \offprints{diego.goncalves@unicsul.br}

\institute{N\' ucleo de Astrof\' isica Te\' orica, CETEC - Universidade Cruzeiro do 
Sul, Rua Galv\~ ao Bueno 868, CEP 01506-000 S\~ao Paulo - Brazil             
		}

   \date{received  ; accepted }

 
  \abstract
   {The dust component of the interstellar medium (ISM) has been extensively studied in 
the past decades. 
Late-type stars have been assumed as the main source of dust to the ISM, 
but recent observations show that supernova remnants may play a role on the ISM dust 
feedback.}  	
   {In this work, we study the importance of low and high mass stars, as well as their 
evolutionary phase, on the ISM dust feedback process. We also determine the changes on the obtained 
results considering different mass distribution functions and star formation history.}         
   {We describe a semi-empirical calculation of the relative importance of each star at 
each evolutionary phase in the dust ejection to the ISM. We compare the obtained results 
for two stellar mass distribution functions, the classic Salpeter initial mass function and 
the present day mass function. We used the evolutionary track models for each stellar 
mass, and the empirical mass-loss rates and dust-to-gas ratio. }  
   {We show that the relative contribution of each stellar mass depends on the used 
distribution. Ejecta from massive stars represent the most important 
objects for the ISM dust replenishment using the Salpeter IMF. On the other hand, for the 
present day mass function low and intermediate mass stars are dominant.}  
   {We confirm that late-type giant and supergiant 
stars dominate the ISM dust feedback in our actual Galaxy, but this may 
not the case of galaxies experiencing high star formation rates, or at high redshifts. In those cases, SNe are dominant in the 
dust feedback process.}


   \maketitle
%

\section{Introduction}

RGB and AGB stars are known as the major continuous dust producers in the Universe, but 
also SN remnants have shown fast grain growth and dusty shells of 
M$_d$ $\sim 10^{-2} - 10^{1}$ M$_{\odot}$ \citep{noz03}. 
From the classical nucleation theory, the timescale for grain growth in the ISM can be estimated by
$\tau_g = 4 s a (f n_i m_i v_i)^{-1}$, where $a$ is the dust mean size, $s$ is the material density, $f$ 
the sticking probability, 
$n_i$ the gas phase density, $m_i$ the atom mass and $v_i$ the velocity of the i-th atom to be added 
onto the grain surface. Considering typical ISM parameters, $a \sim 0.1 - 1 \ \mu$m, $s \sim 2$ g cm$^{-3}$, 
$n_i \sim 1$ cm$^{-3}$, 
$T \sim 10$ K, and assuming an efficiency $f = 0.1 - 1$, we find $\tau_g < 10^9$ yr. 

Dust particles are likely to be destroyed by shock waves \citep{draine79, mckee89}.
Grain-grain or ion-grain collisions will lead to the shattering process, reducing or destroying dust particles. 
\citet{jones94, jones96} described the shattering process of ISM dust induced by SN blasts, obtaining destruction 
timescales of $\tau_{d} \sim 4 - 6 \times 10^8$ yr. Actually, 
as mentioned in these works, the accurate derivation of destruction timescales depends on the velocity of the 
shock waves, the frequency of SNe and the physical properties 
of the ISM, as density and temperature. 
Hence, the destruction timescales are shorter than the dust growth scale,  
the observed dust could not be explained by nucleation in loco, 
but had to be recently ($< 10^9$ yr) injected into the ISM \citep{tielens98}. 

It is commonly suggested that the ISM dust should be mostly originated from evolved low and intermediate mass stars but 
recent observations showed the presence of large quantities of dust 
($M_d = 10^8$ M$_{\odot}$) in the early Universe (z $> 5$) \citep{hughes98,arc01,dunne03,bert03,maio04}. 
At that age low and 
intermediate mass stars were not 
evolved yet and, therefore, SNe are recognized as   
responsible for such material ejections \citep{tod01,mor03,sug06,dwek07}. 

In the post-shock phase of the SN remnant, the gas is generally cool and dense enough to allow dust formation and 
growth \citep{fal03,fal05}.  
Observationally, it is confirmed for 
SN1987A, which shows a $10^{-3}$ M$_{\odot}$ dust shell with a dust to gas ratio for heavier elements $\sim 0.3$. 
 The same result was obtained for several other galactic and extragalactic SNe \citep{bar05,gomez07}. 
Therefore, SNe seem to play an important part on the ISM dust replenishment process \citep{dwek98}. 
However, if the short dust destruction timescale is taken into account, SNe could only be the main 
source of dust of the Galaxy if it presented a high star formation rates in its recent history. 

In this work we present a semi-empirical model, in which we study the role of different 
stellar mass functions in the output of dust ejected to the ISM. The model is described in Sect. 2, in Sect. 3 we 
show the main results and present a brief discussion, followed by the conclusions.

\section{The Model}

To obtain the total amount of dust ejected by a certain star during all its evolution we have to integrate the 
dust mass loss rates of each evolutionary phase over the evolutionary time. The difficulties on performing such 
calculation lie on the determination of the dust mass loss rates at each evolutionary phase of the star, as well as 
its duration. To accomplish this we used the evolutionary tracks given by \cite{sch92}, 
which take into account the mass loss during the stellar evolution. These numerical calculations also provided 
the abundance of heavier elements on the stellar surface at each epoch of the stellar evolution. 
We used these numerical results to determine the amount of material able to be added into the dust grains that is being 
ejected from the stars. 

The total dust mass ejected by each star over its lifetime can be determined by integrating the mass loss rate at 
each stellar evolutionary phase, determined by:

\begin{equation}
\label{eq1}
\int^{M_2}_{M_1} \int^{t_2}_{t_1} \Phi(m) \dot{M_d}(m)\ dt \ dm, 
\end{equation}

\noindent
where $\Phi(m)$ is the stellar population mass distribution, 
$\dot{M_d}(m)$ is dust mass loss rate of a given star of mass $m$ at the time $t$, M$_1$ and  M$_2$ 
are the limits of the mass range and t$_1$ and  t$_2$  are the limits of the time interval. To compute 
the dust mass loss rates we used:

\begin{equation}
\label{eq2}
\dot{M_d}{\rm (t,m)} = \dot{M}{\rm (t,m)} f_d{\rm (t,m)} \chi{\rm (t,m)}, 
\end{equation}
 
\noindent
where $\dot{M}$ is the total mass loss rate, which is obtained empirically 
as described in the following subsection, $f_d$ is the 
dust-to-gas fraction of elements heavier than He and $\chi$ is the wind metallicity, of an star with mass $m$ 
at an evolutionary time $t$.

In Fig. 1 we illustrate the evolutionary track given by \cite{sch92} for stars with masses ranging from 
0.8 to 120 M$_{\odot}$ used in this work, for an initial metallicity of 0.02, and the empirical 
stellar mass loss rates.

\begin{figure}
\centering
\includegraphics[width=12cm]{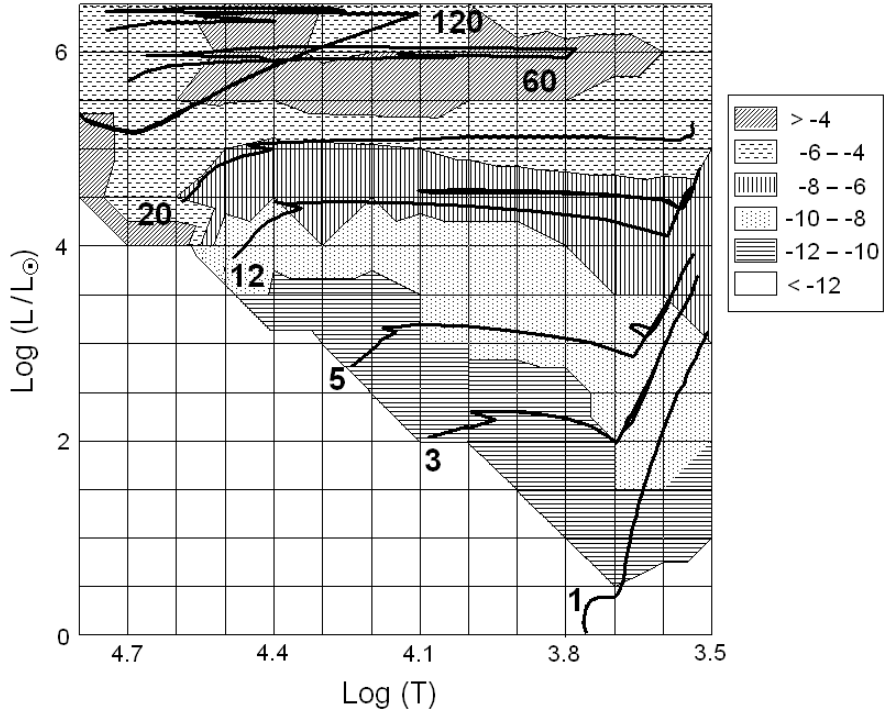}
\caption{Evolutionary tracks for stars with masses ranging from 
1 to 120 M$_{\odot}$ for metallicity of 0.02. The box shows the values of $log(\dot{M})$.}
\label{hrdiagram}
\end{figure}

Different models of homogeneous and heterogeneous condensation theories have been proposed but the actual 
dust-to-gas fraction of 
stellar winds is still not completely understood. The gas acceleration is responsible for the gas rarefaction at 
the base of the stellar atmosphere and, depending on the  
model used, may result in very different pressure and temperature profiles, which are critical to the derivation of the 
condensation rates. As an approximation, we introduce a single dust-to-gas fraction for each stellar mass, depending 
on the chemical composition. The yields for dust-to-metal fraction ($f_d$) for the massive stars were the same used 
by \cite{dwek07} (table 2), and for low and intermediate mass stars we used the data from 
\cite{mor03} (table 1).  

\subsection{The stellar mass-loss rates}

The mass loss rates, for each stellar mass at each evolutionary phase, were introduced empirically. 
For low and intermediate mass stars during the main sequence, giants and supergiants evolutionary phases, 
we used the data from \cite{jag88}. The 
wind velocities for giant and supergiant stars are, in general, $v < 200$ km s$^{-1}$ and 
dust is not destroyed at heliopause. 
At the end of their lives, intermediate mass-stars present very high mass loss rates at the post-AGB's phases 
($\sim 10^{-4}$ M$_{\odot}$ yr$^{-1}$) \citep{wachter02}, and are known as progenitors of the planetary nebulae.

Massive stars are also know to present dust in their winds, mainly at the later stages of 
their evolution. WR and LBV stars present high mass loss rates in unstable and clumpy winds that favor dust 
nucleation and survival. 
For these objects we used the data given by \cite{crow97}. 
Also, about $30 \% $ of these objects are in binary systems, and \cite{march02} 
found that binary systems present high dust production rates due to the 
wind-wind shock, with $4 - 6 \% $ of the wind mass condensed into grains at 
WR+OB systems. 
On the other limit, low mass protostars were also considered 
since their jets and disks are dust growth sites. Typically, these objects present a 
mass-loss rate of $10^{-8}$ M$_{\odot}$ yr$^{-1}$ during $10^6 - 10^7$ yr 
\citep{mundt87}.

\subsection{The stellar mass distributions}

To compute the total contribution for the ejected material by all stars we may 
define a mass distribution function. To simulate the distribution of a 
young stellar population we used the typical Salpeter single component IMF \citep{sal55}: 
\begin{equation}
\label{eq3}
\Psi \left( \log M\right) =AM^{-x}
\end{equation}
\noindent 
for M$_*>1$M$_\odot$, where $A=4.43\times 10^{-2}$ and $x=1.3$, and:
\begin{equation}
\label{eq4} 
\Psi \left( \log M\right) = B\exp \left[ -\left( \log M-\log M_{c}\right)^{2}/2\sigma ^{2}\right] 
\end{equation}
\noindent
for 
M$_*<1$M$_\odot$, where $B = 0.158$, $M_{c} = 0.079$ and $\sigma = 0.69$. For the old populations, which exhibit a significantly lower number of 
massive stars, we used the Present Day Mass Function (PDMF). For the PDMF Eqs. (\ref{eq3}) and (\ref{eq4}) are still valid 
but, $A=4.4\times 10^{-2}$ and $x=4.37$ for $0\leq \log M\leq 0.54$, $A=1.5\times 10^{-2}$ and $x=3.53$ for 
$0.54<\log M\leq 1.26$ and $A=2.5\times 10^{-2}$ and $x=2.11$ for $\log M>1.26$ \citep{chab03,elme04}. 
 
\section{Results and Discussion}

If one assumes that the dust is rapidly destroyed in the ISM ($\tau_{d} \sim 10^8$yr), it is possible to naively 
separate in two the possible 
scenarios of the galactic history: i- old feedback process, which occurred 
in the presence of a large number of massive stars and, ii- 
recent feedback process, with a relatively larger number of 
low and intermediate mass stars. Therefore, we must study the dependence of different stellar mass distributions on 
the dust feedback process.
We performed the calculation of Eqs. (\ref{eq1}) and (\ref{eq2}) computing the ejected material during the evolution 
of all stars using the IMF and PDMF distributions. 

\begin{figure}
\centering
\includegraphics[width=12cm]{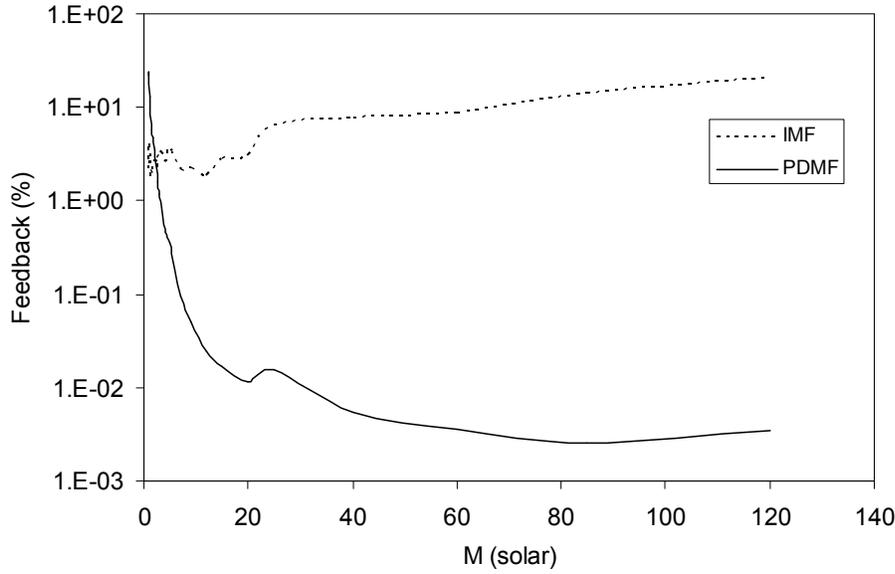}
\caption 
{Dust mass, relative to total (in percent), ejected to the ISM by each stellar 
mass range during its complete evolution. The solid line 
represents the model using a typical Salpeter IMF, and the dotted line 
represents the model for the present day mass function (PDMF).}
\label{dustmass}
\end{figure}

In Fig. 2, we show the relative amount of dust ejected by each stellar mass range 
to the ISM during all its evolution, for the IMF and the PDMF.
It is noticeable that the high mass component is the major source of dust for the Salpeter IMF distribution. 
These stars evolves rapidly and eject large amounts 
of heavy atoms during the SN phase, and their remnants evolve to form shells of few solar masses 
of dust particles. On the other hand, the result is the opposite if use the present 
stellar distribution of the Galaxy. There are few high mass stars, and as a consequence SNe are not so frequent. 
Using the PDMF, the low and intermediate mass stars, as they leave the main sequence, are the main contributors of 
dust to the ISM. 
Intermediate mass stars at supergiant phase have been assumed to be the most important source of dust but, surprisingly, 
the results 
for the PDMF show that stars with $M_* < 3$ M$_{\odot}$  are dominant. It can be explained by the overabundance 
of low mass stars in this distribution. 

In the present stage of the Galaxy as the dust is 
destroyed, or severely changed, in short timescales ($\tau_{d} \sim 10^8$ yr ), the PDMF would 
give more accurate abundances of dust in the ISM. Since the dust 
generated from high mass stars in the past was recycled, and with the absence of 
these objects in the current population, we may conclude that low mass stars are the main source 
of the dust observed in the present stage of the Galaxy.

In Table 1 we show the quantitative results of the calculations described above. For a typical 
IMF the high mass stars ($M_* > 8$ M$_{\odot}$) are responsible 
for 68\% of the dust that returns to the ISM. On the other hand, considering a PDMF, this contribution 
falls to less than 1\%. To determine the absolute feedback mass, we used a total stellar population of 
$1 \times 10^{10}$ M$_{\odot}$, which gives a total dust mass $M_{d} \sim 7 \times 10^{7}$ for the IMF distribution, and 
$2 \times 10^{7}$ for the PDMF.

\begin{table}
\label{dust}
\begin{minipage}{0.5\textwidth}
\caption{\small Relative and absolute dust ISM feedback contribution  
 for different stellar component mass functions.}
\centering
\begin{tabular}{ccc}
\hline\hline
Stellar mass range & IMF & PDMF \\ 
\hline
$M_{d}$ ($M>8$ M$_{\odot}$) & 68\% & 1\% \\ 
$M_{d}$ ($M<8$ M$_{\odot}$) & 32\% & 99\% \\ 
$M_{d\_total}$ \footnote{ {\scriptsize assuming a total stellar population of $1 \times 10^{10}$ M$_{\odot}$.}} 
($M_{\odot}$) & $7 \times 10^{7}$ & $2 \times 10^{7}$ \\
\hline
\end{tabular}
\end{minipage}
\end{table} 

Interestingly, the results indicate that in recent starburst regions, one should expect a larger 
dust-to-gas ratio when compared to an evolved population. It could possibly be the reason for irregular and 
spiral galaxies, 
which present high star formation activity, have more dust than the evolved elliptical galaxies 
\citep{seaq04}. However, since we have very different scenarios here, with different feedback and 
destruction timescales for each type of galaxy, 
this statement needs more detailed calculations to be tested.

\subsection{Time evolution}

In the previous calculations, to obtain the total amount of dust ejected to the ISM we 
had to assume the mass function of the stellar population, which translates the evolutionary phase 
of the stellar component at the epoch we are studying as the dust is recycled in 
short timescales. However, in order to obtain the time evolution of the ISM dust component, we have to include in 
Eq. (\ref{eq1}) the star formation rate function over history. 

We performed the evolutionary calculation assuming a constant star formation rate of 
5 M$_{\odot}$yr$^{-1}$, as used by \cite{mor03}. However, differently of that work we took into account the 
dust destruction and studied its role on the total dust ejected to the ISM. At each time step, we calculate the 
total ejection from each stellar mass range of the current population, add new stars using the 
Salpeter IMF, remove stars that have already evolved and, finally remove the destroyed dust from the total solid component. 
In the present calculations, in order to simulate the chemical 
enrichment of the stars, a metallicity of $z=0.01z_{\odot}$ was arbitrarily used for the initial population 
($t < 10^7$yr), and $z=z_{\odot}$ for $t > 10^6$yr. In Fig. 3 we show the results for the absolute dust 
mass ejected to the ISM. 

\begin{figure}
\centering
\includegraphics[width=12cm]{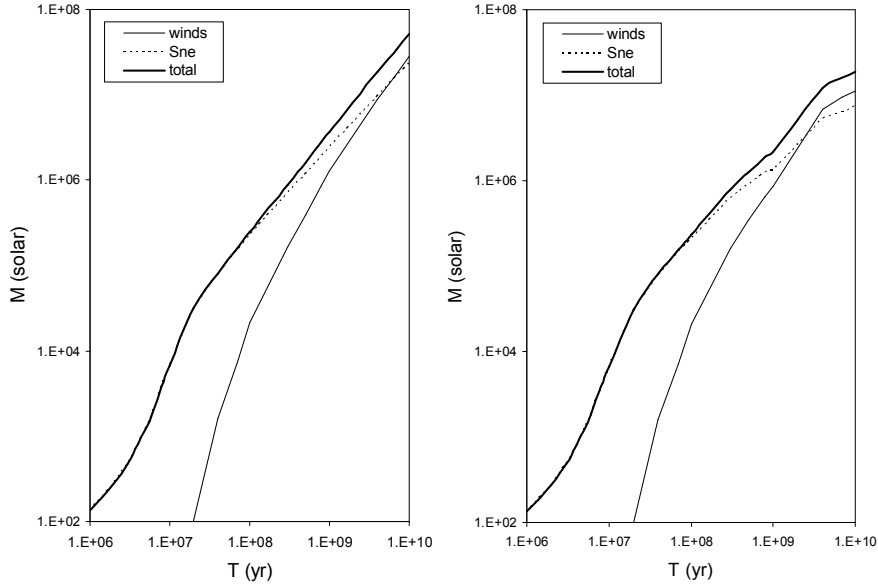}
\caption 
{Absolute dust mass ejected to the ISM. The modeled ejection by high mass stars (solid thin), 
low and intermediate mass stars (dashed) and total (solid thick), without dust 
destruction (left panel) and considering dust destruction (right panel).}
\label{dustage}
\end{figure}

In both cases we can identify the dominance of massive stars on the total ejections at the earlier stages of the 
galaxy evolution. On both models, it is also noticeable the appearance of substantial dust amount from low and intermediate 
mass stars only after $\sim 10^8$ yr. These objects become the major dust producers on the last 0.5 billion years on 
both models. The main differences appear on the later stages, where for considering dust destruction, massive stars 
would be responsible for 3 times less dust mass than obtained by previous calculations, while this proportion for 
low and intermediate mass stars is negligible. After 10 billion years of evolution, we obtained a current dust mass  
of $\sim 6\times10^7$ M$_{\odot}$ without dust destruction, and $\sim 10^7$ M$_{\odot}$ taking into account 
the dust destruction. 

\section{Conclusions}

It is still unclear what is the main source of the dust 
feedback in our Galaxy. The dust must be formed in stars and ejected 
to the ISM and some authors have argued favoring 
cool late-type stellar winds, which present high mass-loss rates and dust is 
proved to be formed in these sites by observations. On the other hand, other plausible 
sources are the SNe ejecta. During the final evolutionary phases, high mass stars 
explode and supersonically eject a very rich gas to the ISM. 

Firstly, to determine the relative amount of 
dust ejected to the ISM for each stellar mass at each evolutionary phase 
we calculated the dust ejection during each evolutionary phase of the stars for different 
stellar mass distributions. 
As main result we showed that SNe are the main source of ISM dust feedback if a classic 
Salpeter IMF distribution is assumed. On the other hand, if we use the present 
day mass function, we show that the main sources of dust to the present 
Galaxy are the low and intermediate mass stars, representing more than 90\% of 
the total dust mass.

Secondly, we studied the dust feedback process along the galactic time evolution, as 
done by \cite{mor03}, but including the effects of dust destruction by SN blasts. 
For simplicity we used a constant destruction rate, consistent with 
the current galactic physical parameters. During previous ages the 
SNe ejecta are dominant, in agreement with previous works. We showed that, considering the dust destruction, low and 
intermediate mass stars are dominant for a galactic age of $t > 10^9$yr, in a much higher proportion. The 
total dust mass of $\sim 10^7$ M$_{\odot}$ is obtained for a star formation rate of 5 M$_{\odot}$yr$^{-1}$.

The dust destruction timescale depends on the SNe frequency, as well as the ISM density and temperature. It 
is probable that the destruction 
rate was higher earlier during the galactic evolution. In this case, the presented conclusions will stand, and 
the role of low and intermediate mass stars 
in later stages will be even higher.

\begin{acknowledgements}

D.F.G thanks FAPESP (No. 06/57824-1) for financial support, also A. Jones and the referee for the comments and suggestions. 
  
\end{acknowledgements}

\end{document}